# Community Time-Activity Trajectory Modelling based on Markov Chain Simulation and Dirichlet Regression

Chen Xia[1], Yuqing Hu[2], Jianli Chen[3]

1Department of Architectural Engineering, The Pennsylvania State University, University Park, PA 16802, United States of America; e-mail: cpx5037@psu.edu
2Department of Architectural Engineering, The Pennsylvania State University, University Park, PA 16802, United States of America; e-mail: yfh5204@psu.edu
3Department of Civil and Environmental Engineering, University of Utah, Salt Lake City, UT 84112, United States of America; email: jianli.chen@utah.edu

*Corresponding email:* yfh5204@psu.edu

## Abstract

Accurate modeling of human time-activity trajectory is essential to support community resilience and emergency response strategies such as daily energy planning and urban seismic vulnerability assessment. However, existing modeling of time-activity trajectory is only driven by socio-demographic information with identical activity trajectories shared among the same group of people and neglects the influence of the environment. To further improve human time-activity trajectory modeling, this paper constructs community time-activity trajectory and analyzes how social-demographic and built environment influence people's activity trajectory based on Markov Chains and Dirichlet Regression. We use the New York area as a case study and gather data from American Time Use Survey, Policy Map, and the New York City Energy & Water Performance Map to evaluate the proposed method. To validate the regression model, Box's M Test and T-test are performed with 80% data training the model and the left 20% as the test sample. The modeling results align well with the actual human behavior trajectories, demonstrating the effectiveness of the proposed method. It also shows that both social-demographic and built environment factors will significantly impact a community's time-activity trajectory. Specifically: 1) Diversity and median age both have a significant influence on the proportion of time people assign to education activity. 2) Transportation condition affects people's activity trajectory in the way that longer commute time decreases the proportion of biological activity (eg. sleeping and eating) and increases people's working time. 3) Residential density affects almost all activities with a significant p-value for all biological needs, household management, working, education, and personal preference.

# 1. Introduction

People's time-activity trajectory is defined as a sequence of activities performed by individuals at various places during the twenty-four hours of day and night to satisfy biological needs, institutional, personal obligations, and personal preferences (Hägerstrand, 1974; Drummond, 1995). While the goal of urban management is driven by a desire to improve human life in settlements (Mattingly, 1994), sudden events such as natural disasters, power outages, fire hazards, and man-made events often threaten the fulfillment of those needs, disrupting normal routines and household comfort. On some occasions, people's time-activity trajectory impacts the vulnerability of people exposed to these events and contributes to how the community responds and assigns resources when unexpected events hit (Debnath, 2013). For example, as an important clue of personal exposure to air pollution, the time-activity trajectory is often incorporated into the stochastic exposure model to evaluate how urban environment contaminants influence citizen life (Matz, et al, 2014). On the other hand, people's activity trajectories are also closely related to energy consumption in buildings, especially in the residential sector, and thus play an important role in community energy pattern analysis (Karatasou et al, 2013). Due to a lack of detailed activity trajectory data, current community energy analysis often uses simplified predefined occupancy values to run the model, which leads to discrepancies between the simulated energy shape and the actual consumed energy. However, for occupancy schedules, studies have pointed out that people in different communities hold different schedules, and these differences can greatly influence the final building stock energy simulation results compared with those using national average schedules (Buttitta et al., 2019). This makes accurate modeling of people's time-activity trajectory crucial to support community resilience management and emergency response strategies.

Many technologies and methods have been developed to capture time-activity trajectories, such as Geographic Information Systems (GIS), social media data mining, and time use survey (Kwan and Neutens 2014; Siła-Nowicka et al, 2016; Allahviranloo et al., 2017; Huang,2021). Time-use surveys often collect data from computer-assisted telephone interview (CATI) technology to record individual activity in 24 hours. For example, Large-scale time-activity pattern data have been collected in North America, including the Canadian Human Activity Pattern Survey (CHAPS) (Matz et al, 2014), the National Human Activity Pattern Survey (NHAPS) (Klepeis et al, 2001), and American Time Use Survey Data (Hamermesh et al, 2005) conducted in the US. Geographic Information Systems (GIS), and social media data mining collect data from handheld GPS (Global Positioning System) units, GPS-enabled smartphone tracking applications, or social media such as Twitter (Qi & Du, 2013).

Compared with survey data, modern technology makes it possible to automatically capture geographic locations in real-time at detailed spatial scales through digital devices either embedded in vehicles or carried by people on smartphones (Kwan and Neutens 2014). However, some limitations exist, such as data with only location and time cannot report context and cause activity ambiguities (Miller, 2021). In more recent years, social media mining is adopted to help gain the exact activity people are doing. However, data collected in this social media-based method only reflect people who used the application at a specific time (Lu, 2021). In addition, it also can cause representativeness bias for people who are less likely to use social media (e.g., children and the elderly), while they can be relatively vulnerable groups. Therefore, many studies built their time-activity trajectories model based on people's time use survey data, which can compensate for limitations in social media-based tracking methods to some degree (Allahviranloo et al., 2017; Hafezi et al, 2021).

Time-activity trajectory simulation models can be categorized into the mechanism-identified activity model and data-driven model. Mechanism-identified activity models examine the underlying behavioral mechanisms based on theories like utility maximation and decision-making theory (Kitamura, 2000). This approach investigates how people assign time to various activities based on individual characteristics, environmental constraints, and personal/household utility (Ellegard & Vilhelmson,2004; Zhang et al, 2002). Compared with the mechanism-identified model which generates the time-activity trajectory through analyzing the human decision-making process, the data-driven model often uses statistics to generate a time-activity trajectory at population based on the large-scale monitoring time use data (Hafezi et al, 2021). However, most research using this method focus on how social-demographic features influence people's time-activity trajectories and tend to ignore the built environment. Under this method, trajectories are generated through clustering with the assumption that people in the same group keep similar trajectories. After groups are formed, researchers can evaluate the influence of social demographic by analyzing the attributes of people in different groups. However, people with the same social attributes may still perform different activities at the same time because of various built environments. For example, people working in the same company may leave and get home at different times due to traffic conditions around in their respective residential neighborhoods.

To fill these research gaps, we propose a methodology to model community time-activity trajectory with the proportional distribution of people performing different activities in each time step and analyze how socio-demographic and built environments influence the trajectories. This contains two steps: the first is to use the widely used Markov chains to generate community trajectories based on the American Time Use Survey (ATUS). This method has been developed and widely used in many studies. However, it highly relies on the survey data, which does not cover all communities and only has small data samples in some communities that are not enough to represent the community trajectory. In this study, we apply this method to construct the original trajectories for those communities with data. The second step is to analyze how the community demographic attributes and built environment influence the communities' trajectories and try to model the trajectories based on these attributes with the proposed Dirichlet regression model.

The paper is organized as follows: The literature review part provides an overview of time-activity trajectory models and points out gaps in current modeling methods. The methodology section presents the Markov Chain-based activity modeling method and the Dirichlet Regression to analyze how social and built environments influence the activity trajectory. The Results & Discussion section presents the findings of the case study on New York City and provides model validation results. Finally, the conclusion section highlights the contributions of the study on community time-activity trajectory modeling and activity influences analysis and the limitations and prospects for future study direction.

## 2. Literature Review

While the time-activity trajectory represents people's spatiotemporal activities that occurred over time, in this study for the purpose at hand we assume that we are dealing with an urban region and 24 hours. Since the interest in urban management is driven by a desire to improve human life in settlements (Mattingly, 1994), people's time-activity trajectory constitutes an important part of urban-related research. For example, as an important clue of personal exposure to air pollution, the time-activity trajectory is often incorporated into the stochastic exposure model to evaluate how urban environment contaminants influence citizen life (Matz, et al, 2014). Human time-activity is also closely related to energy consumption in buildings, especially in the residential sector, and thus plays an important role in energy demand simulation models (Karatasou et al, 2013), which will impact community vulnerability when expose to power outage. Due to these applications, the modeling of people's time-activity trajectory is attracting increasing attention.

A wide array of theories and methods have been developed to generate people's time-activity trajectory. One basic stream is to examine the underlying mechanisms of how people assign time to various activities. The mechanisms identified can either be based on constraints or from the point of human utility (Janssens, et al 2004). The core idea of the constraints-based model is that individuals face many constraints limiting their choices. Typically, there are three types of constraints, i.e., (1) capability constraints that limit the activities of an individual for biological reasons, such as the necessity of sleeping a minimum number of hours and the intervals of eating regularly; (2) authority constraints that generally refer to the legal environment rules such as access time restrictions to different places; (3) coupling rules that define where, when, and for how long different people can meet for a joint activity (Rasouli and Timmermans, 2014). Utility-based models assume that individual or household allocates time to activities based on utility-maximizing theory (Zhang et al, 2002). This approach uses utility maximization-based equations to identify the relationships between an individual's characteristics and their activity choices.

Mechanism-identified activity models establish the time-activity trajectory based on analyzing factors on the individual's choice-making, which can consider influences from the aspects of both individual and environment. However, some limitations have been pointed out by previous studies (Daisy et al, 2018). One is that they assume individuals are rational without considering uncertainty in the model, which makes it difficult to determine the statistical significance of the factors affecting the individual's decisions on their activity. The other is that most decision-making models depend on predefined parameters and fail to update with dynamic data. In this case, the data-driven model can serve as a solution, which utilizes the monitoring data provided by time-use surveys or GIS sensors to generate people's time-activity trajectories. During the modeling process, data mining techniques are often used to cluster groups with common activity patterns, then various statistical techniques are employed to explore the influence of socio-demographic features on human activity trajectories. For instance, Jiang et al (2012) used the principal component analysis (PCA) and K-means clustering algorithm to cluster several representative groups according to people's activities and then compared the social demographic differences in each cluster. Stating that traditional principal component analysis only represents frequent activities and disregard the infrequent ones, Liu et al (2015) developed a novel process derived from Hidden Markov Models (pHMMs) to quantify the occurrence probabilities and sequence of all daily activities. To address the uncertainty in start time and activity durations, Hafezi et al (2021) formulate a Random-Forest model based on people's socio-demographic characteristics and temporal features of their activities to help predict activity patterns with specific start time and activity duration.

Compared with models figuring out specific activity decision mechanisms, a data-driven model can quickly update according to real-time data while reducing uncertainty underlying the human decision-making process with large-scale data. However, existing studies often ignore the built environmental influences. For example, in Allahviranloo et al (2017) 's study on community activity profiles, they used a K-mean clustering algorithm to identify representative activity pattern clusters and then captured the correlation among individual demographic profiles and the activity sequences using multivariate probit models. When doing clustering with similar activity patterns, humans are separated from the environment they locate. After clusters are formed, the influence of social-demographic can be analyzed by studying the features of individuals inside different groups. The influences of the built environment are often left out in current research.

The built environment is defined as human-made surroundings that provide a setting for human activity, ranging in scale from personal shelter to neighborhoods and large-scale civic surroundings (Butt et al, 2015). Its influence on people's activity has been studied from two aspects: the permanent built environment where people reside and the context changed with people's location mobility. For the context changes with people's mobility, as mechanism-identified activity models point out, the environment is an important constraint on people's time-activity decisions. Empirical evidence suggests that activities are highly related to built environments. For example, some building types play significant roles in people's daily life. They carry a particular semantic meaning such as the living and working places, the restaurant and shopping mall, etc (Hillsdon et al, 2015). For the built environment where people reside, it has been stated for years that the design of the community's built environment influences some of human's specific activities, such as health-related activity (Kerr et al, 2012) and criminal activity (MacDonald, 2015). However, limited research analyzes how the community's built environment influenced people's daily activity trajectories based on openly accessible data. Therefore, in this study, we use data-driven methods to generate people's time-activity trajectories, and then further analyze how social and built environments influence these trajectories based on Dirichlet regression analysis.

## 3. Methodology

### 3.1 Approach Overview

The common definition of a community is a group of people with diverse characteristics who are linked by social ties, share common perspectives, and engage in joint action in geographical locations or settings (Al, 2001). According to the various research purpose, the range of the geographical scale can be from small census block groups to county, city, or even countrywide (Chi, 2012). In the United States, a county is an administrative or political subdivision of a state that consists of a geographic region with specific boundaries. Considering the small census block groups generally contains between 250-550 housing units, the built environment between nearby block groups does not show much difference. So, this study focus on a county-level community trajectories simulation.

To construct a time-activity trajectory that could be used to analyze both social-demographic and built environment. As Figure 1 shows, we first use Markov chains to generate time-activity trajectories for different areas. Then, we gather built and socio-environment information from different areas and use Dirichlet Regression to analyze how these social and built environments influence the activity trajectory distribution. Finally, we apply this regression model to perform activity trajectory prediction based on social and built environment information. To validate the model, we use New York City as a case study and gather data from American Time Use Survey, Policy Map, and the New York City Energy & Water Performance Map to test the proposed method. During the prediction process, we use 80% data to train the regression model and the left 20% as the test sample. In this study, each dependent variable of the regression model is a compositional distribution of the probability of different activity categories, which are rarely analyzed with the usual multivariate statistical methods and do not have a standard to perform model verification. In this paper, Box's M Test and T-test are performed for the equality test of covariance structure and center of test sample and modeling result. Detailed contents of each step are further classified in the following part.

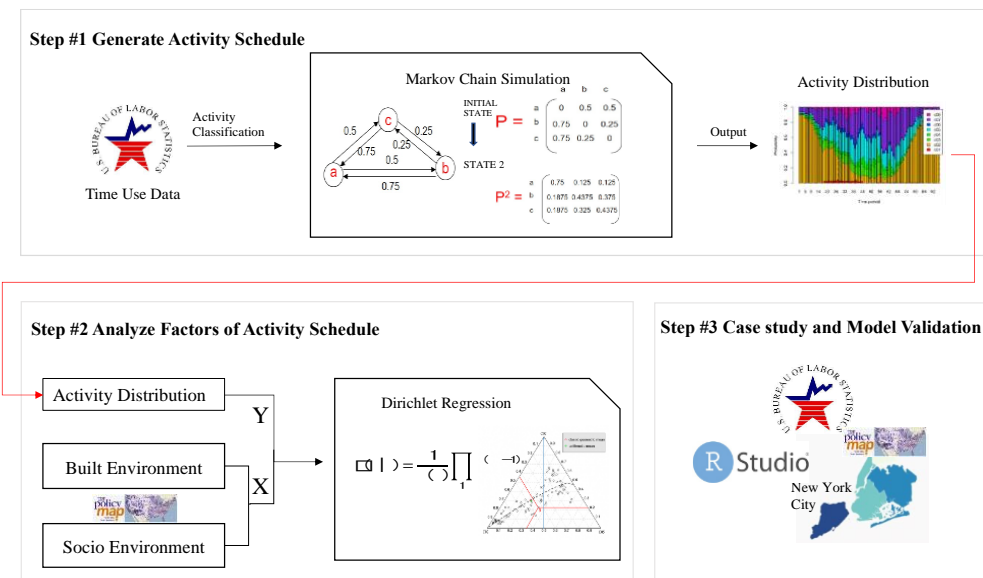

Figure 1. Community time-activity trajectory modeling framework

## 3.2 Markov Chains Simulation for Community Time-activity trajectory Modeling

The term 'activity' refers to various behaviors over time (Drummond, 1995). There is no standard uniform classification of personal activities yet. The most accepted typology is in travel behavior studies proposed by Reichman, which divides activities into three categories as subsistence, maintenance, and discretionary (leisure) activities (Reichman, 1976; Chung and Lee.2017). Yamamoto and Kitamura (1999) advocate a simplified activity classification of two categories: mandatory (must-engaged activity) and discretionary (individual has the choice to be engaged). Other classifications like physiological needs, institutional demands, personal obligations, and personal preferences are also employed in research (Vilhelmson, 1999). This study builds activities categories based on the original activity classes in the American Time Use Survey, which category activities into 18 types. Because some of its activity categories serve the same purpose according to the standards in the literature, we further summarized detailed activity categories like socializing, relaxing, and sports into the personal preference category used in the literature. Finally, we have 8 classifications as follows: essential health activity, biological needs (eating, sleeping), working, education, household management, personal obligations (shopping, banking, childcare, etc.), personal preference (leisure activities), and others (outside traveling activities).

Basic time-activity trajectory refers to the fundamental time geographic entity representing the ordered mobility of individuals moving in geo-space during a day cycle (Frihida et al, 2004). For a more complex level, not only the geographic location but also the activity performed at specific times and locations are also included. Data used to construct time-activity trajectory can be collected through time-use surveys, Geographic Information Systems (GIS), and social media data mining. Since this study focuses on synthesizing chains of activities and space-time distribution at the community level based on information regarding all-day activity categories, social demographics of individuals, and the building environment they locate, survey data is preferred as a source for constructing people's time-activity trajectory.

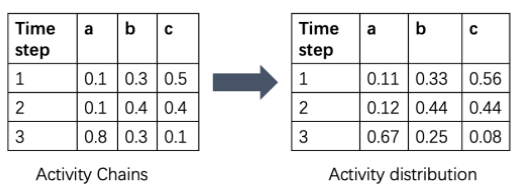

Figure 2. Time-activity pattern transformation example

To generate activity schedules in each community, one way is to directly obtain it from the sampled time-use data. However, this method is deterministic in a certain sense that only collected individual activity schedules can be generated, which is insufficient to fully describe the stochastic nature of individual behaviors. Since we focus on NY as a specific study region, the number of collected individual behavior schedules usable for behavior sampling and analysis is limited. This would potentially introduce bias or incomprehensiveness in analysis. Hence, in this study, we use Markov Chain (MC) based model, as a stochastic modeling approach, to simulate the activity status of individuals in each time step. This method has been widely used in activity simulation research because of its advantage to handle categorical time-series sequences (Liu et al, 2015; Zhou et al 2020) as well as capturing behavior stochasticity. Markov Chain assumes that the probability of observation in any state is influenced only by the preceding states (Geyer, 1992). When the simulation reaches a certain scale, it can include more diverse sets of behaviors while ensuring the convergence of behavior probability profiles with stochasticity considered. The used MC approach makes the analyzed behavior profiles closer to the actual human behaviors.

During the modeling process, it captures the temporal characteristics in two aspects: transition probability between different activities and the duration of each activity. In this study, we divide 24h into 96-time states. For a discrete sequence of states, given the first-order Markov chain with an initial probability α and a transition matrix $\xi_{pq}$, the probability of transitions from State p to State q in Sequence $i$ could be calculated with the stationary transition matrix $\xi^i_{pq}$. In this study, the transition probability matrices are trained based on existing samples. The Bayes-based approaches can be further utilized to update transition probabilities (Li&Ji, 2020; Chen et al, 2019) with new coming data.

Once we define an initial activity in the start time step (e.g., Sleeping at 11 pm), we can get the probability of each activity in the following time step sequences.). As Figure 2 shows, for activity A, B, C, they have separate occurrence probability for daily time steps t (1,2,3…t) as A ($a_1$, $a_2$, $a_3$…$a_t$), B ($b_1$, $b_2$, $b_3$…$b_t$), C ($c_1$, $c_2$, $c_3$…$c_t$). To transfer them to the broken activity probability distribution in the same step t, there should be probabilities: $a_t/(a_t + b_t + c_t)$; $b_t/(a_t + b_t + c_t)$; $c_t/(a_t + b_t + c_t)$. For this study, the time-activity trajectory generated will be a 96*8-dimension matrix.

### 3.3 Dirichlet Regression Model for Activity Influences Analysis

Activity trajectories generated in Markov Chain simulation are groups of compositional data of activity possibility distribution in each time step. Compositional data are a special case in the field of statistics. They are non-negative data with the distinctive property that the sum of their value is a constant, usually 1 (proportions) or 100% (percentages) (Aitchison, 1994). Compared with the general data set, applying the usual multivariate analysis for such compositions can lead to problems ignoring their underlying sample constraints (strictly positive and sum to a constant value). To study this type of data, there is a way to do log ratios of the compositional data, so the traditional multivariate techniques can be applied to the transformed data (Aitchison,1994)). After that, Campbell and Mosimann developed an alternative approach by extending the Dirichlet distribution to a class of Dirichlet Covariate Models (Dirichlet Regression), which can be used to analyze a set of variables lying in a bounded interval without having to transform the data (Hijazi, 2009). Therefore, to analyze how social and built environments influence activity possibility distribution in each time step of activity trajectories and mode the compositional data in each step, the Dirichlet regression model was proposed in this study.

Dirichlet distribution is a family of continuous multivariate probability distributions parameterized by a vector α of positive reals (Maier, 2014). As defined in Equation 1, let Y = ($y_1$,…,$y_D$) be a 1 x D positive vector having the Dirichlet distribution. Each variable is shapely parameterized by the vector $α_d$, which has the same number of elements as our Y vector. For this distribution, the parameters $a_d > \forall D$, $y_D \in (0,1)$, $and \sum_{D=1}^{D} y_D = 1 \forall D$ must hold.

$$f(y|a) = \frac{1}{B(a)} \prod_{1}^{D} y_D^{(a_D-1)} \tag{1}$$

In the Dirichlet regression model, $y \sim f(a)$ denotes a variable that is Dirichlet - distributed with the common parametrization. For those variables, the sum of all as- --$α_0 = \sum_{D=1}^{D} a_D$--- can be interpreted as a 'precision' parameter. The higher this value, the more density is near the expected value). The expected values are defined as E[$y_D$] =$a_c/a_0$. Parameter estimation in the Dirichlet regression model is also based on maximum likelihood techniques (Hijazi, 2006). Once the estimation has been accomplished, the fitness of the Dirichlet model to the compositional data needs to be assessed. Since the likelihood-based methods depend on the parametric assumption led to inaccurate results, it is important to check the validity of the parametric assumption. Widely used techniques in regression model assessment such as examination of the residuals, goodness-of-fit measures, and influence diagnostics are also used in this study to investigate the goodness of fit of the estimated Dirichlet models, and the importance of factors included.

For independent variables included in the Dirichlet regression model, potential factors of social and built environments are summarized from the literature. Social environments have been researched soundly in current research, either in the mechanism-identified modeling or data-driven model, which mainly includes demographic information such as age, gender, and education (Matz, et al, 2015; Dianat et al, 2020). Therefore, this study uses these people's socio-demographic information to represent the social environments. The built environment generally includes broad indicators in terms of both indoor and outsides conditions. Studies on built-environment indicators in the United States show most indicators fall into three domains: land use, housing condition, and transportation (Lynch and Mosbah, 2017). While housing condition is often linked to specific house unit and is more related to household activity, previous studies pointed out that land use and transportation are two important factors that influence human activity trajectories (Bhat et al, 2013). Therefore, this study also takes the two factors. For the transportation environment, while peak-hour traffic often reflects the maximum capacity and availability of urban traffic (Winston, 1991), the average travel time to work, a key index defined by the US census bureau to represent the transportation infrastructure availability in the community, is used as an indicator. As for the land use, we use the building type, which is defined based on occupancy use in the International Building Code (IBC), as the indicator.

To validate the model prediction result, part of the samples are used as training data and the left as testing data. Since compositional data are rarely analyzed with the usual multivariate statistical methods, Dirichlet regression as a recently proposed method to model does not have a standard to perform model verification. In this paper, we check whether there is a significant difference between model prediction results and original data to verify the model. To compare two groups of compositional data, Pawlowsky et al (2007) proposed that there is no difference between both groups if the covariance structure and the centers are all the same. Box's M Test is used to perform the covariance structure test. The null hypothesis for this test is that the observed covariance matrices for the dependent variables are equal across groups. Therefore, a non-significant test result indicates that the covariance matrices are equal. A T-test is used to perform a center test. The null hypothesis for the t-test is also there is no significant difference between the two groups.

## 4  Case study

We test this method in New York City areas since all its five counties have very different environments and data on both activities and influences are available in this city. For the activity data, we use the American Time Use Survey (ATUS), a publicly available continuous survey on people's time use in the United State, to perform the activity simulation process. The major purpose of ATUS is to develop nationally representative estimates of how people spend their time. To reach this goal, individuals are generally interviewed through computer-assisted telephone interview (CATI) technology about how they spent their time on the previous day, where they were, and whom they were with. This survey provides information on the time people spend in around three hundred activities, such as sleeping, eating, socializing, and relaxing. Besides activities, demographic information, such as gender, race, age, occupation, income, marital status, and the presence of children in the household, also is available for each respondent. The survey is performed nationwide for a continuous year since 2003. In this study, we collected part of the survey data from 2015 to 2019 in New York City in weekdays, including a total of 1284 individuals. Then we derived the city into five communities according to its county separation, which are New York County (Manhattan), Kings County (Brooklyn), Bronx County (The Bronx), Richmond County (Staten Island), and Queens County (Queens) to analyze their built-environment impact.

**4.1 Data process**

Socio-demographic information in each community is collected from Policy Map, an online mapping platform (www.policymap.com) created by the Reinvestment Fund featuring data on demographics, housing, mortgage loans, and FEMA (Federal Emergency Management Agency) disaster declarations. As Table 1 shows, we collected nine socio-demographic characteristics in the 2015-2019 period. Some information is recorded as the data is, like median age, male to female ratio, people with disabilities, household median income, and annual unemployment rate. Others are collected through specific indexes defined in Policy Map Data Dictionary: population density is the estimated number of people per square mile; diversity is collected through an index ranging from 0 to 87.5 that represents the probability that two individuals of different races are chosen at random in the given geography; Racial segregation is collected through Theil Index, which is a measure of how evenly members of racial and ethnic groups are distributed within a region, calculated by comparing the diversity of all sub-regions (Census blocks) to the region as a whole; Education is estimated percent of people with at least a high school diploma.

Table 1 Socio-Demographic Information in Communities

| Social environment | Bronx | Queen | Manhattan | Kings | Richmond |
|---|---|---|---|---|---|
| Population density | 34,090.12 | 21075.68 | 71,488.69 | 36,573.40 | 8,135.86 |
| Diversity | 59.34 | 76.39 | 68.31 | 72.57 | 56.54 |
| Racial segregation | 0.26 | 0.36 | 0.33 | 0.43 | 0.29 |
| Median Age | 34 | 39 | 38 | 35 | 40 |
| Male to female ratio (%) | 89 | 94 | 90 | 90 | 94 |
| People with disabilities (%) | 15.23 | 9.61 | 10.28 | 9.98 | 9.83 |
| Household income median ($) | 40088 | 68666 | 86553 | 60231 | 82783 |
| Annual unemployment rate (%) | 5.3 | 3.4 | 3.4 | 4 | 3.8 |
| Education (%) | 72.76 | 82.02 | 87.28 | 82.38 | 88.75 |
| Number of Samples | 102 | 116 | 124 | 151 | 25 |

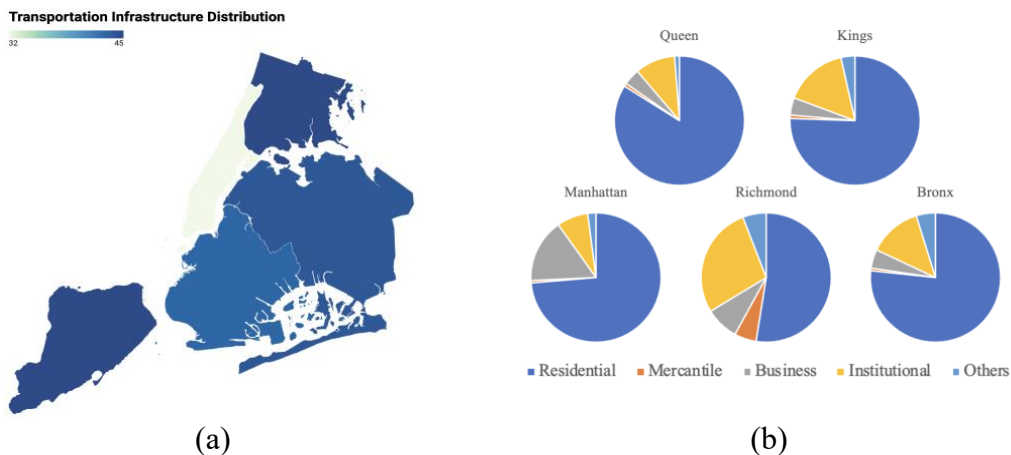

(a) (b)

Figure 3. Transportation infrastructure availability (a) and
building type distribution (b) in New York

Building type and transportation are two important aspects of the built environment reported in the literature that will influence human activities. Figure 3(a) shows the distribution of average travel time to work in New York City. Among the five boroughs, work commuting time in Manhattan is obvious the least than in the other four communities. This indicates workers living in principal cities of micropolitan areas generally have a shorter average travel time than those living elsewhere in micropolitan areas. Building type distribution information is gathered from the NYC (New York City) Energy & Water Performance Map developed by the New York University's Marron Institute of Urban Management and the NYU Urban Intelligence Lab in partnership with the Mayor's Office of Sustainability. As Figure 3(b) shows, residential building occupies most of all of the communities. While institutional building comes second, in the Richmond community, it is highest. Of the business building, most are located mostly in Manhattan County, though it is geographically the smallest of the five boroughs of New York City.

Activity data are integrated into three distinct groups in each community through K-means clustering according to their different distributions on active at home, sleep, and out of the home. For each cluster in Table 2, the X-axis stands for a scaled time through a day from 0 h to 24h, the Y-axis stands for data points. Different colors in visualization denote the different activity status of each data point in each period. Overall, there exist two obvious patterns among these clusters: most time at home and most time away from home. For each community, the differences are in two aspects: one is data distribution in clusters. Take Richmond and Manhattan as an example, most data samples are in Cluster 1 as active at home during the daytime in Richmond, while in Manhattan, most people are away from home during the daytime at Cluster 2. The other is that though two clusters are all most of the time away from home, the structures inside clusters are different, which reflects when people begin to transfer from one state to the other. Take Cluster 3 in Bronx and Queen as an example, people in the Bronx tend to stay at home for some time and leave home at noon, while people in Queen tend to leave home early and spend more time active at home in the evening.

To make sure that the regression model is a valid one, a correlation test was performed to deal with potential multicollinearity issues among independent variables. In general, the more predictor variables included in a valid model the lower the bias of the predictions, but the higher the variance, and tend to be over-fitting. Coefficients in Table 3 shows there is a high correlation between some variables like Diversity and Racial Segmentation (0.81). For these highly related variables whose coefficients are larger than 0.75, we do filter with the Akaike information criterion (AIC). Given a collection of models for the data, AIC and BIC estimate the quality of each model. BIC is more useful in selecting a correct model while the AIC is more appropriate in finding the best model for predicting future observations (Chakrabarti & Ghosh, 2011). Generally, the smaller AIC & BIC value, the better the model. For prediction, AIC was preferred when contraction existed in this study. After this process, variables left in the model are Time, Diversity, Med_Age, Residential, and Trans, among which time is the predictor variable while other variables are regarded as moderating variables.

Table 2. Activity cluster in New York counties

| County | Cluster 1 | Data | Cluster 2 | Data | Cluster 3 | data |
|---|---|---|---|---|---|---|
| Bronx | 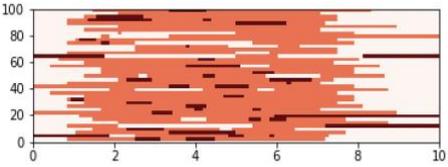 | 39% | 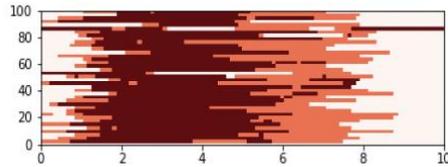 | 38% | 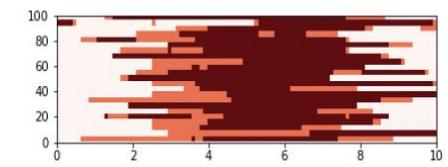 | 23% |
| Queen | 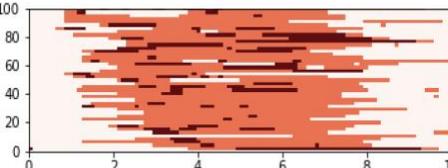 | 38% | 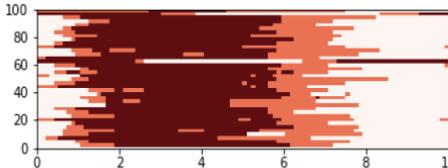 | 33% | 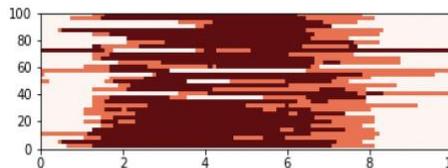 | 29% |
| King | 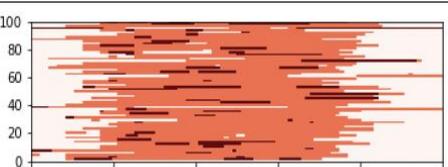 | 42% | 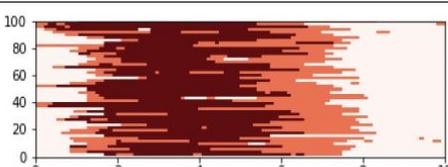 | 33% | 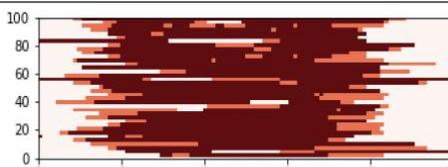 | 25% |
| Richmond | 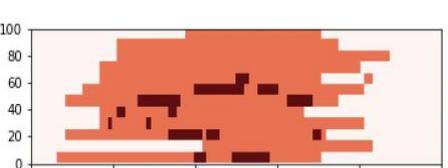 | 48% | 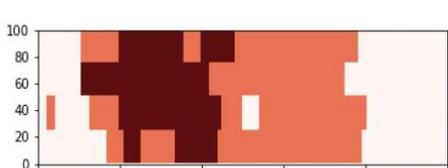 | 16% | 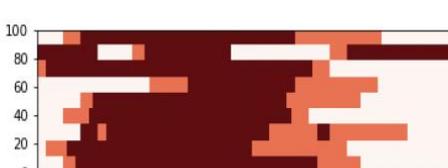 | 36% |
| Manhattan | 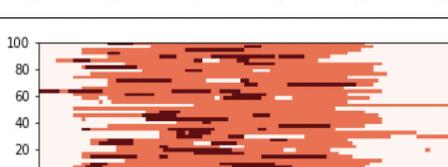 | 31% | 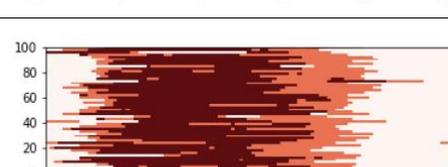 | 44% | 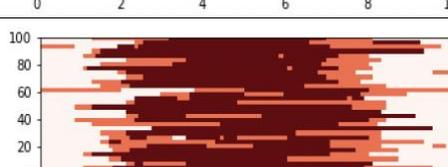 | 26% |

X-axis: Time; Y-axis: Data sample; Color: ▢ Sleep; ▬ Active at home; ▬ Out of home

## 4.2 Results & Discussion
**Community time-activity trajectory patterns**

Time-activity trajectory patterns in each community are generated through Markov Chain. Figure 5(a) shows probability distribution through a single day (24h) in one community. The X-axis denotes every 15 minutes as one time step, while the y-axis denotes the probability of each activity in that time step. An obvious trend in the figure is that during the night, biological activity(c02) occupies the most possibilities. Though there is still some amount of people awake for personal preference at night, there is nearly no probability for other activity. When time shifts to daytime, other activities began to grow in possibility, especially the activity of working. The working probabilities surges around time step 20 at 5 am. If we see the mean probability in Figure5(b), biological activity(c02), working(c05), and personal preference(c07) dominants people's daily life, which fits with common sense.

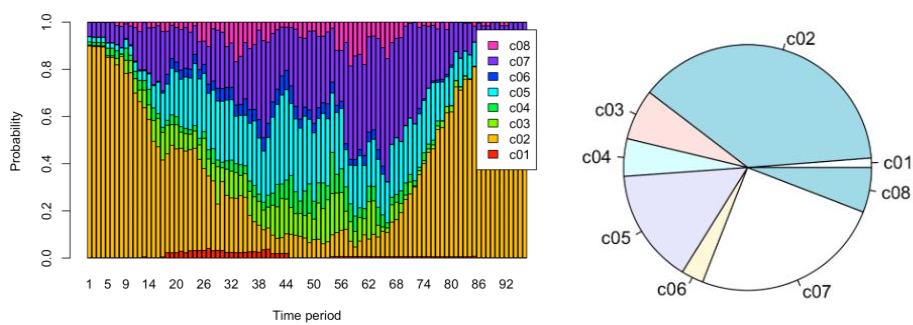

Figure 5. (a) Time-activity trajectory patterns through 24h (left);
(b) Mean activity occurrence probability (right)
(Note: c01- health emergency, c02-biological needs, c03- household management, c04-personal obligation, c05-working, c06-education, c07-personal preference, c08-others)

For probability distribution of specific activity during a day, biological and working activity trajectories in different communities are compared in Figure 6. Since the X-axis denotes each 15 minutes time step and starts from midnight, communities share a similar trajectory in the night with a very low probability of working. During the daytime, we see that some community like the Bronx maintains a relatively stable trajectory with low probability varying between 0.2 and 0.3. However, in other community, especially Manhattan, the working probability surges around period 20 at 5 am and go a sharp down around period 65 at 4 pm. And it hits the highest working probability during the day. If we focus on the working lasting period, though probability shows relatively low in the Bronx, it lasts longest till around 7 pm. While in other communities, the working trajectory follows a regular working period. If we compare the biological pattern in the two communities with the most different working trajectories, though in general, these two trajectories follow the same time pattern, biological activity in Manhattan is ahead by around one hour more than activity in the Bronx. The difference could be caused by both social and built environment differences such as population and working structure, which will be explored in the following section through the Dirichlet regression model.

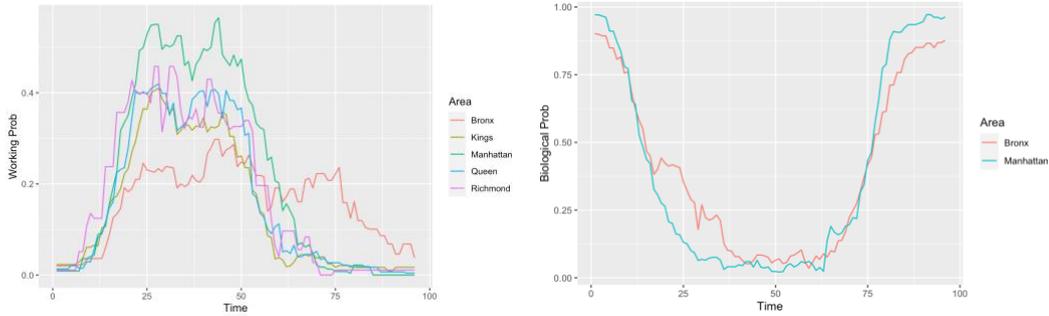

Figure 6 Biological and working activity trajectories in different communities.

The detailed activity probability distribution in each time step can be analyzed through ternary diagrams, a widely used technique to visualize compositional data. It can only display three parts at the same time, so the three main activity biologicals, working, and personal preference are displayed in Figure 7. While each angle demotes the activity with 100% probability, the perpendicular to the opposite side acts as an axis, on which the intersection denotes 0 probability. Take Point D as an example, it has a 0.35 probability for activity c05, 0.43 probability for activity c02, 0.22 probability for activity c07. The dotted line in this figure is the mean proportion of each activity with time varies, accuracy parameter controls how data is displayed around the mean. During the modeling process of Dirichlet regression, the mean and the accuracy parameter are estimated. In this study, variation of activity shows a strong relationship with time. So, in the modeling process, time was regarded as the predictor variable while the other environment variables as moderating variables, which are variables that affect the direction and/or the strength of the relationship between predictor and outcome variables (Thompson, 2006).

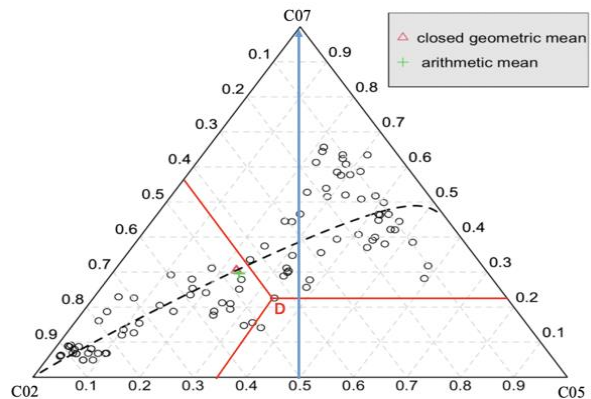

Figure 7. Ternary diagrams of activity composition

**Regression modeling results and discussion**

Dirichlet regression results are shown in Table 4. Time as the direct predictor variable shows significance for all activities. In this study, instead of using time, time square is adopted as the predicted variable because the visualization of activity sequences suggests that there is possible a non-linear relationship with this variable. So, the coefficient of the time variable determines how wide or narrow the related time-activity graphs are, and whether the graph turns upward or downward. A positive value causes the ends of the parabola to point upward, while the negative caused the other direction. Results show that it is negative for all activity but biological, which makes sense as the main biological activity is sleeping, which reaches the highest proportion at night, while other activity shows the least proportion. The values also control the rate of the change. The greater the quadratic coefficient, the narrower the parabola. The lesser the quadratic coefficient, the wider the parabola. Which may cause different time-activity trajectories in various communities. This Difference in values of different areas may be caused by the moderating effect of the social and built environment.

For the social environment effect, diversity has a significant positive effect on biological activity and education. This indicates that with diversity increases, the portion of people performing these activities will increase at a specific time, especially for the education activity, which holds the largest coefficient with the most significant p-value. For diversity, low diversity index values suggest more homogeneity and higher index values suggest more heterogeneity. Racially and ethnically homogenous areas are sometimes representative of concentrated poverty or concentrated wealth (Parker, 2015). High heterogeneity, therefore, contributes to more diverse activity patterns. Besides, studies on the relationship between diversity and education reveal that racial diversity could promote education for individuals, institutions, and the whole society (Milem, 2003). The result in this study also reflects one aspect of the diversity benefits from the perspective of the proportion of educated people. Median age gives a picture of what the age distribution looks like in a specific community. In this study, median age shows a negative effect on education, which indicates that with the aging of a specific population, people spend less time studying. This makes sense as education is generally performed in the early years of people's life. As people get aged, though there are still opportunities for people to get educated, it is less common than that at their young period.

For the built environment, transportation influences both biological activity and working activity, though in opposite direction. Since average travel time to work was used as an index for transportation conditions. The negative effect on biological shows that with the increase of traffic time to work, people spend less time on biological activity, in which sleeping may occupy the most part. This result is consistent with the 2019 American Community Survey Reports (Burd, 2021). In their survey, they found that departure for the workplace holds a relationship with traffic time. Generally, workers having the longest average travel times to work leave home during the earliest hours of the day. For example, Workers leaving home during the earliest hours of the day from 12:00 a.m. to 4:59 a.m had the longest average travel times to work at 35.2 minutes, while workers who left for work from 6:00 a.m. to 8:29 a.m reported the longest average travel time to work at 32.8 minutes. This also partly explains why average travel time to work has a positive effect on working time. To reduce traffic, some people may leave home earlier and stay office later (Hartgen et al, 2012).

The proportion of residential buildings affects almost all activities with a significant p-value for all biological needs, household management, working, education, and personal preference. While high residential building proportion often points to being a more urbanized area, studies have provided evidence that rural residents experience a health disadvantage compared to urban residents due to their different daily activity patterns (Matz et al, 2014). Besides, the positive effect on personal preference indicates that as residential building increases, people have more time for personal preference activities, such as relaxing and doing sports. Studies on the residential environment reveal that higher residential building density has direct effects on utilitarian physical activity and has many benefits in terms of efficient use of infrastructure, housing affordability, energy efficiency, and possibly vibrant street life (Forsyth et al, 2007), which somewhat contributes to the result that people in higher residential building density area spend less time on household activity or working activity.

### 4.3 Model prediction and verification

To verify the model, 80 percent of samples are used as training data and the left 20 percent as testing data. Box's M test and T-test are used to check the equality of covariance structure and the centers respectively. For the covariance structure test, the null hypothesis is that observed covariance matrices for the dependent variables are equal across groups. Therefore, a non-significant result indicates there is no significant difference between the two groups, which is verified in the results shown in Table 5. T-test for center equality has a similar hypothesis as Box's M test. However, for the test result category centers in each group, activity emergency health and education show a difference with significant p values. Analyzing the data, the absolute difference between the centers of these two activities are smaller than other activity. the main reason for the significant test results lies in that the proportion of these two activities is far smaller than other activity, which magnifies the difference.

Table 5 Model verification results

|  | C01 | C02 | C03 | C04 | C05 | C06 | C07 | C08 |
|---|---|---|---|---|---|---|---|---|
| Box'M test | Chi-Squared = 3886.1, df = 4656, p-value = 1 | | | | | | | |
| T test | | | | | | | | |
| p-value | 3.6e-12 | 0.5742 | 0.50724 | 0.0323 | 0.2483 | 1.1e-6 | 0.1126 | 0.2130 |
| Mean (Test) | 0.0073 | 0.3487 | 0.0635 | 0.0414 | 0.1873 | 0.0115 | 0.2088 | 0.0494 |
| Mean (Predict) | 0.0183 | 0.3748 | 0.0678 | 0.0051 | 0.1661 | 0.0233 | 0.2398 | 0.5764 |

# 5. Conclusion

In this paper, we present and test a method modeling community detailed time-activity trajectory patterns consisting of both social demographic and build environment. we start with generating different time-activity trajectory patterns for different communities by using Markov chains. To construct the activity trajectory, we divide one day into 96 time periods with 15 minutes intervals. For each activity, Markov chains were used to estimate activity chains for 24 hours. Then, all activity chains were integrated and transformed into a probability distribution of activities in each period. With this compositional trajectory result, a prediction model based on Dirichlet regression was built for the future prediction of community time-activity trajectory prediction. During the modeling process, the correlation among community environments to the bundles of activities performed and their corresponding time sequence has also been captured. To verify the model prediction accuracy, we apply the model in New York City as a case study with 80 percent of samples as training data and the left 20 percent as testing data. The predicted result of test data is the same as the sample in terms of the covariance matrix and center for most activity in two groups' activity trajectories, which proves the robustness of the proposed method in modeling community time-activity trajectory in detail distribution.

This paper also quantifies the influence of different environment variables on the distribution of community activities proportion in terms of time. While time is the main influence for community time activity trajectory pattern, both social-demographic and build environment could influence activity proportion's changes over time. The result shows that 1) Diversity and median age both have a significant influence on education activity, though in a different direction. 2) Transportation condition affects people's activity trajectory in the way that longer commute time decreases the proportion of biological activity and increases people's working time. 3) Residential building proportion affects almost all activities with a significant p-value for all biological needs, household management, working, education, and personal preference. Since current models developed to generate time-activity trajectories are often based on pure socio-demographic information with the assumption that people in the same group follow the same trajectory, the methodology presented in this paper could model community time-activity trajectory with a detailed distribution of people performing different activities in a specific period considering both socio-demographic and built environment influences. This could help provide a more profound and accurate basis for the study that is based on human activity trajectories.

The proposed community time-activity trajectory model can also help improve the community's resilience against sudden events, such as natural disasters, power outages, fire hazards, and man-made events by understanding its initial impact. The in-time activity distribution could help related organizations evaluate potential damage severity and make in-time response strategies. Besides that, this can also benefit daily energy planning to improve the community's resilience. For example, the daily load management and scheduling, especially in high-demand times under extreme hot and cold days. If we know the in-time community activity distributions among areas in different communities, we can make daily load management and scheduling with minimum economic and safety risks. Also, while social demographic information cannot be changed, the revealed influences on human activity of built environment and infrastructure could be used by the governor to promote community wellbeing.

However, there are still some limitations to this study. At present, the transition probabilities matrix in this study is generated based on existing samples. To appropriately incooperate uncertainty when the model is built with small sample size in specific communities, the transition probability matrix can be updated when additional records come in. Also, since there is no standardized way to validate the Dirichlet regression model with compositional data, we use Box's M test and T-test to compare the equality of covariance structure and center to test the equality of two group data. Additionally, due to some activity features and the limitation of telephone-based survey limitation, some acclivity occupies a small proportion, which makes it hard to perform accurate predictions on those activities. However, these activities, especially health emergency activities, are often related to vulnerable people who are most likely to be affected during the power outage period. So, in the future, if the information on these activities could be gathered through other sources like social media or hospital data platforms and integrated into the activity simulation model, the research will be more complete. Besides that, while the unit community selected in this paper is at the county level, the method can be applied to communities with the bigger geographic area such as cities or states as the data used is openly accessible. When expanding the geographic area, more built environment factors can be explored besides the typical two built environment attributes we focused on in this study. In the time dimension, as we modeled the community time-activity trajectories on weekdays, weekend patterns should also be considered. However, compared with weekday trajectories, people's activities tend to be more random on weekend days, so, this may rely on in-time GIS tracking technology and can be studied in the future.

Table 3 variables Correlation Matrix

|  | PD | Diversity | Racial_seg | Med_Age | M_F_Ratio | Disabilities | Hou_Minc | Unemp | Edu | Trans | Int | Res | Mer | Bus |
|---|---|---|---|---|---|---|---|---|---|---|---|---|---|---|
| **Population density** | 1.00 | 0.26 | 0.14 | -0.29 | -0.69 | 0.08 | 0.16 | -0.13 | 0.04 | -0.90 | 0.38 | 0.57 | -0.72 | 0.65 |
| **Diversity** | 0.26 | 1.00 | 0.81 | 0.01 | 0.07 | -0.49 | 0.07 | -0.54 | 0.05 | -0.23 | 0.34 | 0.48 | -0.58 | -0.02 |
| **Racial_seg** | 0.14 | 0.81 | 1.00 | -0.08 | 0.00 | -0.63 | 0.12 | -0.49 | 0.24 | -0.11 | 0.18 | 0.23 | -0.32 | -0.09 |
| **Med_Age** | -0.29 | 0.01 | -0.08 | 1.00 | 0.87 | -0.71 | 0.85 | -0.78 | 0.80 | -0.14 | 0.76 | -0.76 | 0.65 | 0.49 |
| **M_F_ratio** | -0.69 | 0.07 | 0.00 | 0.87 | 1.00 | -0.63 | 0.51 | -0.58 | 0.53 | 0.34 | 0.40 | -0.73 | 0.71 | -0.01 |
| **Disabilities** | 0.08 | -0.49 | -0.63 | -0.71 | -0.63 | 1.00 | -0.79 | 0.94 | -0.86 | 0.24 | -0.73 | 0.49 | -0.32 | -0.38 |
| **Hou_Medi_Income** | 0.16 | 0.07 | 0.12 | 0.85 | 0.51 | -0.79 | 1.00 | -0.87 | 0.97 | -0.56 | 0.92 | -0.62 | 0.43 | 0.81 |
| **Annual_unemp** | -0.13 | -0.54 | -0.49 | -0.78 | -0.58 | 0.94 | -0.87 | 1.00 | -0.84 | 0.47 | -0.91 | 0.36 | -0.17 | -0.59 |
| **Education** | 0.04 | 0.05 | 0.24 | 0.80 | 0.53 | -0.86 | 0.97 | -0.84 | 1.00 | -0.43 | 0.81 | -0.71 | 0.52 | 0.68 |
| **Transportation** | -0.90 | -0.23 | -0.11 | -0.14 | 0.34 | 0.24 | -0.56 | 0.47 | -0.43 | 1.00 | -0.72 | -0.21 | 0.42 | -0.90 |
| **Institutional** | 0.38 | 0.34 | 0.18 | 0.76 | 0.40 | -0.73 | 0.92 | -0.91 | 0.81 | -0.72 | 1.00 | -0.30 | 0.09 | 0.85 |
| **Residential** | 0.57 | 0.48 | 0.23 | -0.76 | -0.73 | 0.49 | -0.62 | 0.36 | -0.71 | -0.21 | -0.30 | 1.00 | -0.97 | -0.20 |
| **Mercantile** | -0.72 | -0.58 | -0.32 | 0.65 | 0.71 | -0.32 | 0.43 | -0.17 | 0.52 | 0.42 | 0.09 | -0.97 | 1.00 | 0.00 |
| **Business** | 0.65 | -0.02 | -0.09 | 0.49 | -0.01 | -0.38 | 0.81 | -0.59 | 0.68 | -0.90 | 0.85 | -0.20 | 0.00 | 1.00 |

Table 4 Dirichlet Regression Results

|  | C01 | | C02 | | C03 | | C04 | | C05 | | C06 | | C07 | | C08 | |
|---|---|---|---|---|---|---|---|---|---|---|---|---|---|---|---|---|
|  | Estimate | Pr | Estimate | Pr | Estimate | Pr | Estimate | Pr | Estimate | Pr | Estimate | Pr | Estimate | Pr | Estimate | Pr |
| **Intercept** | -0.8623 | *** | 0.3345 | *** | 0.7421 | *** | 0.4401 | *** | 1.6749 | *** | -0.4812 | *** | 1.7993 | *** | 0.5554 | *** |
| **Time^2** | -0.1853 | ** | 1.6110 | *** | -0.6456 | *** | -0.5734 | *** | -0.6598 | *** | -0.3473 | *** | -0.4113 | *** | -0.6582 | *** |
| **Diversity** | -0.0224 | 0.8025 | 0.1783 | ** | 0.1126 | 0.1656 | -0.1434 | 0.2894 | 0.1274 | 0.1449 | 0.3271 | *** | -0.0078 | 0.9132 | 0.2102 | * |
| **Med_Age** | -0.0879 | 0.5200 | -0.1210 | 0.2159 | -0.1763 | 0.1393 | 0.1417 | 0.7278 | -0.2325 | 0.0811 | -0.3946 | ** | 0.1722 | 0.1026 | -0.2141 | 0.0834 |
| **Trans** | -0.0751 | 0.2871 | -0.1262 | * | -0.1197 | 0.0578 | -0.0117 | 0.8539 | 0.3596 | *** | -0.0636 | 0.4003 | 0.0237 | 0.6516 | -0.0086 | 0.8888 |
| **Res** | 0.0654 | 0.6791 | -0.4644 | *** | -0.3727 | ** | -0.0581 | 0.7153 | -0.3138 | * | -0.7864 | *** | 0.3466 | ** | -0.4107 | ** |
| | AIC: -14213 | | | | | | BIC: -13897 | | | | | | Log-likelihood:7186 | | | |

Note: (1)c01- health emergency, c02-biological needs, c03- household management, c04-personal obligation, c05-working, c06-education, c07-personal preference, c08-others: (2) Significance codes: 0 '***' 0.001 '**' 0.01 '*' 0.05